\documentclass[aps, showpacs, showkeys, nofootinbib, floatfix]{revtex4}
\usepackage{}

\usepackage{amssymb}
\usepackage{amsmath}
\usepackage{graphicx}

\begin{document}

\title{Unified Dark Fluid with Fast Transition: Including Entropic Perturbations}

\author{Weiqiang Yang}
\author{Lixin Xu\footnote{lxxu@dlut.edu.cn}}

\affiliation{Institute of Theoretical Physics, School of Physics and Optoelectronic Technology, Dalian University of Technology, Dalian, 116024, P. R. China}

\begin{abstract}
In this paper, we investigate a unified dark fluid model with fast transition and entropic perturbations. An effective sound speed is designated as an additional free model parameter when the entropic perturbations are included, and if the entropic perturbations are zero, the effective sound speed will decrease to the adiabatic sound speed. In order to analyze the viability of the unified model, we calculate the squared Jeans wave number with the entropic perturbations. Furthermore, by using the Markov Chain Monte Carlo method, we perform a global fitting for the unified dark fluid model from the type Ia supernova Union 2.1, baryon acoustic oscillation and the full information of cosmic microwave background measurement given by the WMAP 7-yr data points. The constrained results favor a small effective sound speed. Compared to the $\Lambda$CDM, it is found that the cosmic observations do not favor the phenomenon of fast transition for the unified dark fluid model.
\end{abstract}


\maketitle

\section{Introduction}

Accelerating expansion of the universe has been shown from the type Ia supernova (SNIa) observations \cite{ref:SN-Riess1998,ref:SN-Perlmuter1999} since 1998. It has also been confirmed by the Cosmic Microwave Background (CMB) anisotropy measurement from Wilkinson Microwave Anisotropy Probe (WMAP) \cite{ref:CMB-Spergel2003} and the large scale structure (LSS) from the Sloan Digital Sky Survey (SDSS) \cite{ref:LSS-Pope2004}. To explain the mechanism of the acceleration, theorists introduce an exotic energy component with negative pressure, which is called as dark energy. This idea has greatly inspirited theorists to propose a lot of dark energy models. The simplest but most natural candidate of DE is the cosmological constant $\Lambda$, with the constant equation of state (EoS) $w_{\Lambda}=-1$. This model keeps a good fit with the cosmic observations, but it suffers from the fine-tuning problem and the cosmic coincidence problem. To avoid the issues, the unified dark fluid (UDF) models have been put forward and further studied in Refs. \cite{ref:UDF1,ref:UDF2,ref:UDF3-Xu2012-diffcs2,ref:UDF4,ref:UDF5,ref:UDF6,ref:UDF7,ref:UDF8,ref:UDF9,ref:UDF10,ref:UDF11,ref:UDF12}.
In principle, the EoS can be determined to an integration constant by the adiabatic sound speed. Based on this opinion, the so-called $\Lambda \alpha$CDM or constant adiabatic sound speed (CASS) model was proposed \cite{ref:UDF2,ref:UDF3-Xu2012-diffcs2,ref:UDF10}, where a constant adiabatic sound speed $c^2_{s,ad}=\alpha$ was assumed. Meanwhile, The cases of zero and time variable sound speed were respectively discussed in Ref. \cite{ref:UDF11} and Ref. \cite{ref:UDF12}. Besides, the EoS of the UDF models can be specified as some different forms, such as Chaplygin gas (Cg) model and its generalized versions \cite{ref:UDF4,ref:UDF5,ref:UDF6,ref:UDF7,ref:UDF8,ref:UDF9}. The generalized Cg (gCg) model keeps highly consistent with the cosmic observations and becomes indistinguishable from the $\Lambda$CDM model, one can see Ref. \cite{ref:end-Sandvik2004}.

In order to break away from the degeneracy of the UDF models, a kind of unified model with fast transition in the EoS has been put forward in Refs. \cite{ref:fast1-Bruni2012,ref:fast2-Piattella2010,ref:fast3-Bertacca2011,ref:fast4-Felice2012}. If the transition is slow, the differences between the UDF and $\Lambda$CDM are negligible. However, if the transition is fast, the model shows some different and interesting features. From the point of dynamics, the model with fast transition owns a longer matter dominated era than $\Lambda$CDM, and suddenly enters into a late $\Lambda$CDM-like epoch. Several works have been devoted to study the features of unified model with fast transition, the viability of a UDF model has been analysed by the adiabatic sound speed and the functional form of the squared Jeans wave number in Refs. \cite{ref:fast1-Bruni2012,ref:fast2-Piattella2010}. Theoretical analysis tells us that a viable unified model allows for the phenomenon of fast transition in the evolution of EoS. Additionally, the likelihood analysis on three models with varying EoS has been carried out by the cosmic observations in Ref. \cite{ref:fast4-Felice2012}. The results show that the models are not favored over $\Lambda$CDM by the Akaike information criterion.

In Ref. \cite{ref:fast4-Felice2012}, the models with fast varying EoS have been constrained by the so-called CMB shift parameters, i.e. $R$, $l_a$ and $z_\ast$ which are obtained based on the $\Lambda$CDM model. On the one hand, the values of the CMB shift parameters depend on the $\Lambda$CDM model. Therefore, the so-called circular problem would come out if one uses these derived data points to constrain other cosmological models. In fact, the CMB shift parameters would be different for different cosmological models due to different physics process around the last scattering surface, for example early dark energy model Ref. \cite{ref:earlyDE-Reichardt2012} where the contribution from dark energy may not be neglected due to nontrivial EoS of dark energy. Thus, it is dangerous to use the data points due to much departure from $\Lambda$CDM model. On the other hand, the full CMB data points contain more information than the shift parameters. For instance, at the late epoch when the dark energy is dominated, the gravitational potentials decay and the late integrated Sachs-Wolfe (ISW) arises. It affects to the anisotropy power spectra of CMB at large scale (low $l$ parts) due to the ISW effect which is sensitive to the properties of dark energy. Apparently, the CMB shift parameters do not include this information. One can see Ref. \cite{ref:Rshortcoming-Xu2012} about the advantages of the full CMB information.

To fill out the gap and avoid the so-called circular problem, in this paper we will not use the derived CMB shift parameters but the full information from WMAP 7-yr data points to constrain our unified model with fast transition. One can expect a tight constraint. Furthermore, the entropic perturbations are included in the research process. The reason why we consider entropic perturbations is that the adiabatic sound speed is not enough to characterize the micro-scale properties. So, the effective sound speed as a specified model parameter is introduced. For a generalized dark matter, the effective sound speed was defined in Ref. \cite{ref:cs2define-Hu1998}. When the entropic perturbations vanish, the effective sound speed is reduced to the adiabatic sound speed. In Ref. \cite{ref:cs2study-Pietrobon2008}, the model with a constant adiabatic sound speed including entropic perturbations was studied, where the effective sound speed was fixed $c^2_{s,eff}=\alpha,0,1$. Nevertheless, the effective sound speed should be a free model parameter determined by the cosmic observations , one can see Ref. \cite{ref:cs2studyfree-Xu2012}. Following this paper, we consider the effective sound speed $c^2_{s,eff}$ is a free parameter in the range of $[0,1]$.

The paper is organized as follows. In section II, we propose a UDF model with a fast transition in the EoS. And we analyze the viability of the model from the squared Jeans wave number. In section III, by adopting the Markov Chain Monte Carlo (MCMC) method with the cosmic observational data sets which SNIa, baryon acoustic oscillations (BAO) and the full CMB information of WMAP 7-yr data points, we show the model parameter space. Section IV is the conclusion.

\section{A Unified Dark Fluid Model and its perturbation equations}

We introduce a UDF model based on the hyperbolic tangent function \cite{ref:fast1-Bruni2012}, whose EoS is parameterized as
\begin{eqnarray}
w_u=-\frac{1}{2}\tanh \left(\frac{a-a_t}{\beta} \right)-\frac{1}{2},
\label{eq:EoS}
\end{eqnarray}
where the parameter $a_t$ represents the scale factor corresponding the moment of the transition and the constant $\beta$ indicates the rapidity of the transition. The Friedmann equation for the unified model, which reads,
\begin{eqnarray}
H^2=H^2_0\left\{\Omega_b a^{-3}+\Omega_r a^{-4}+(1-\Omega_b-\Omega_r)exp\left[\int^1_a\frac{3}{a'}(1+w_u)da'\right]\right\},
\label{eq:Friedmann}
\end{eqnarray}
where $\Omega_b$ is the dimensionless energy density of baryon, $\Omega_r=2.49\times10^{-5}h^{-2}$ represents the dimensionless energy density of radiation.

\begin{figure}[!htbp]
\includegraphics[width=12cm,height=8cm]{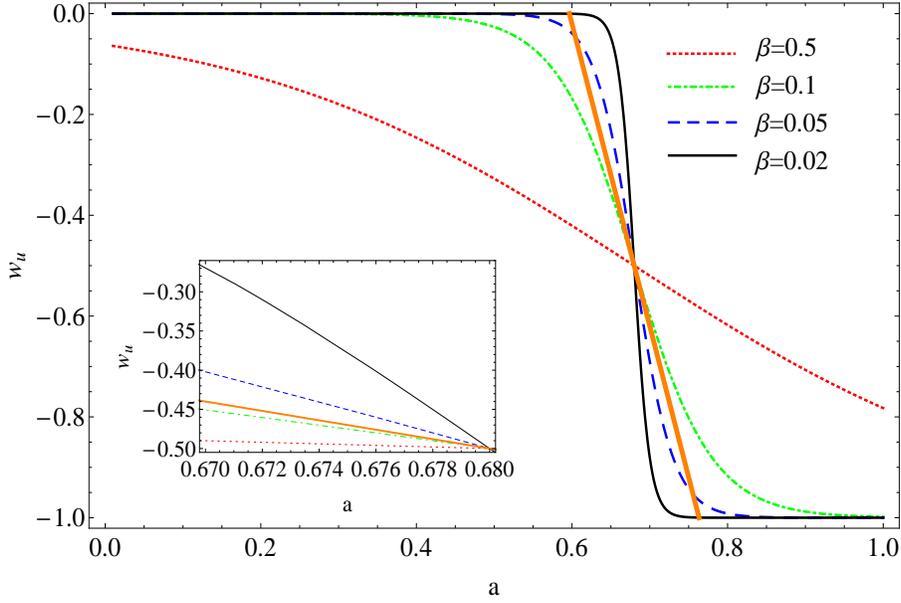}
  \caption{The evolutional curves of EoS for the UDF model when the parameter $a_t=0.68$. The red dotted, green dot-dashed, blue dashed and black solid lines respectively correspond to the case of $\beta=0.5, 0.1, 0.05$ and $0.02$. The thick orange solid line is the critical state between slow and fast transition.}
  \label{fig:wtanh}
\end{figure}

To study the rapidity of the transition about this model, the evolutional curves of EoS have been plotted in Fig. \ref{fig:wtanh} when the model parameter $a_t$ is fixed. When the value of $\beta$ is small enough, the curve around the transition varies rapidly, this phenomenon is the so-called fast transition. Obviously, the smaller the value of $\beta$ is, the faster the rapidity of transition becomes.

Then, we try to define roughly the concept of slow or fast transition from the geometry property of the curve. From the embedded small image in Fig. \ref{fig:wtanh}, fixing on the small interval $[0.67,0.68]$, the critical state between the slow and fast transition is defined as a line (named as CSL) with its slope $k_{CSL} = -6$. Therefore, the slow transition of EoS means that the slope on the small interval is in the range of $(-6,0)$ and the slope of the fast one is less than $-6$. The area above the CSL is the fast transition and the area below the CSL is the slow one. It is easy to see that the transitions corresponding the lines with $\beta=0.5,0.1$ are slow, and the two other cases are fast. Of course, according to our definition, the EoS of $\Lambda$CDM model undergoes a slow transition because its slope is in the range of $(-6,0)$ on the small interval $[0.67,0.68]$.

In the synchronous gauge, the perturbed metric is
\begin{eqnarray}
ds^2=a^2(\eta) \left[-d\eta^2+(\delta_{ij}+h_{ij})(x,\eta)dx^idx^j \right],
\label{eq:permetric}
\end{eqnarray}
where $h_{ij}$ is the metric perturbation and $\eta$ is the conformal time. According to the conservation of the energy-momentum tensor $T^{\mu}_{\nu;\mu}=0$, one can obtain the perturbation equations of density contrast and velocity divergence for the UDF model
\begin{eqnarray}
&&\dot{\delta}_u=-(1+w_u)(\theta_u+\frac{\dot{h}}{2})-3\mathcal{H}(\frac{\delta p_u}{\delta \rho_u}-w_u)\delta_u,
\label{eq:pereq1}
\end{eqnarray}
\begin{eqnarray}
&&\dot{\theta}_u=-\mathcal{H}(1-3c^2_{s,ad})\theta_u+\frac{\delta p_u/\delta \rho_u}{1+w_u}k^2\delta_u-k^2\sigma_u,
\label{eq:pereq2}
\end{eqnarray}
where the dot denotes derivative with respect to the conformal time $\eta$ and the other notations follow Ma and Bertschinger \cite{ref:Ma1995}. And the adiabatic sound speed reads
\begin{eqnarray}
c^2_{s,ad}=w_u-\frac{\dot{w}_u}{3\mathcal{H}(1+w_u)},
\label{eq:cs2ad}
\end{eqnarray}
more details about perturbation theory have been presented in Ref. \cite{ref:per-Hwang2001}. When the EoS of a pure barotropic fluid is negative, it may give rise to the instability of the perturbations, like the w=constant quintessence dark energy model. In order to avoid this issue, we consider a generalized dark matter \cite{ref:cs2define-Hu1998} with permitting the entropic perturbatios and supposing a positive or null effective sound speed, the non adiabatic stress can be separated out as
\begin{eqnarray}
p_u\Gamma_u=\delta p_u-c^2_{s,ad}\delta \rho_u,
\label{eq:enper1}
\end{eqnarray}
which is gauge independent. Using the effective sound speed $c^2_{s,eff}$ in the rest frame of the UDF model, the entropic perturbations become
\begin{eqnarray}
w_u\Gamma_u=(c^2_{s,eff}-c^2_{s,ad})\delta^{rest}_u.
\label{eq:enper2}
\end{eqnarray}

The gauge transformation into an arbitrary gauge
\begin{eqnarray}
\delta^{rest}_u=\delta_u+3\mathcal{H}(1+w_u)\frac{\theta_u}{k^2}
\label{eq:enper-rest}
\end{eqnarray}
gives a gauge-invariant form for the entropic perturbations. Combining Eqs. (\ref{eq:enper1}), (\ref{eq:enper2}) and (\ref{eq:enper-rest}), the perturbation equations can be rewritten as
\begin{eqnarray}
&&\dot{\delta}_u=-(1+w_u)(\theta_u+\frac{\dot{h}}{2})-3\mathcal{H}(c^2_{s,eff}-w_u)\delta_u-9\mathcal {H}^2(c^2_{s,eff}-c^2_{s,ad})(1+w_u)\frac{\theta_u}{k^2},
\label{eq:enpereq1}
\end{eqnarray}
\begin{eqnarray}
&&\dot{\theta}_u=-\mathcal{H}(1-3c^2_{s,eff})\theta_u+\frac{c^2_{s,eff}}{1+w_u}k^2\delta_u-k^2\sigma_u.
\label{eq:enpereq2}
\end{eqnarray}

We use Eqs. (\ref{eq:cs2ad}), (\ref{eq:enper-rest}) and the continuity equation, the perturbation equations become
\begin{eqnarray}
&&\dot{\delta}_u^{rest}=3\mathcal{H}w_u\delta_u^{rest}-(1+w_u)(\theta_u+\frac{\dot{h}}{2})-\frac{9}{2}\mathcal{H}^2(1+w_u)^2\frac{\theta_u}{k^2},
\label{eq:enpereq1R}
\end{eqnarray}
\begin{eqnarray}
&&\dot{\theta}_u=-\mathcal{H}\theta_u+\frac{c^2_{s,eff}}{1+w_u}k^2\delta_u^{rest}.
\label{eq:enpereq2R}
\end{eqnarray}

Combining the above two equations and using the first-order perturbed Einstein equations in Ref. \cite{ref:Ma1995}, the second order differential equation of density perturbation reads
\begin{eqnarray}
\ddot{\delta}_u=\mathcal{H}(6w_u-3c^2_{s,ad}-1)\dot{\delta_u}+[-k^2c^2_{s,eff}+\frac{3}{2}\mathcal{H}^2(1+8w_u-3w_u^2-6c^2_{s,ad})]\delta_u.
\label{eq:2ndper}
\end{eqnarray}
This equation is gauge invariant, one also can see Ref. \cite{ref:jeans-Christopherson2012} where the equation is obtained in the conformal Newtonian gauge. According to the definition of the squared Jeans wave number \cite{ref:absoluteKJ-Bertacca2007}, we can obtain the $k^2_{J}$ for the UDF model with entropic perturbations
\begin{eqnarray}
k^2_J=\frac{3\mathcal {H}^2}{2c^2_{s,eff}}|1+8w_u-3w^2_u-6c^2_{s,ad}|.
\label{eq:jeans}
\end{eqnarray}
As is known, the squared Jeans wave number $k^2_J$ is an important aspect in determining the viability of a unified model because of its effect on the perturbations. In fact, any viable UDF model should satisfy the condition $k^2_J\gg k^2$ for all the scales of cosmological interest, one also can see Refs. \cite{ref:fast1-Bruni2012,ref:fast2-Piattella2010,ref:fast3-Bertacca2011}. Here, we try to find a large enough $k_J^2$ for the model in order to agree with the condition.

\begin{figure}[!htbp]
\includegraphics[width=12cm,height=8cm]{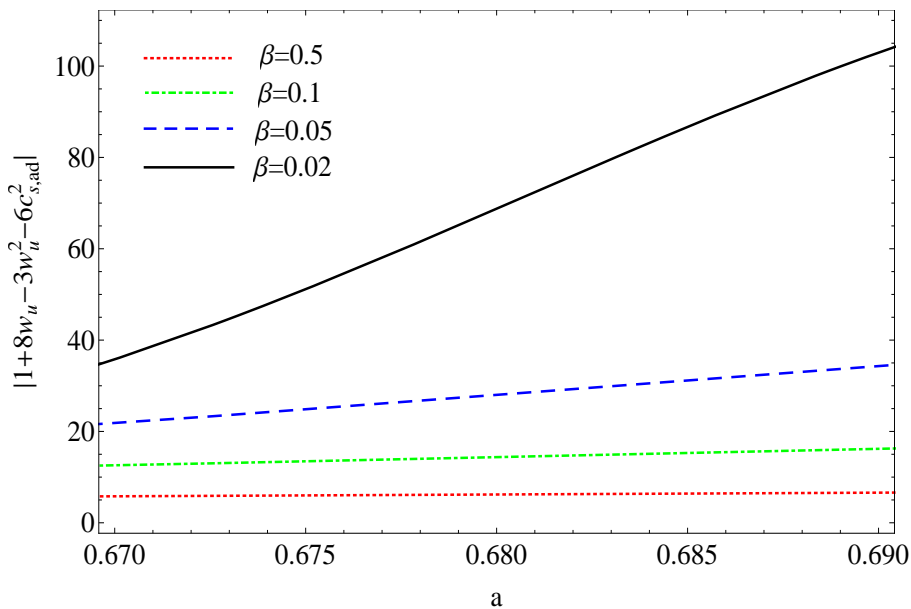}
  \caption{The evolutional curves of $|1+8w_u-3w^2_u-6c^2_{s,ad}|$ for the UDF model on the interval $[0.67,0.69]$ when the parameter $a_t=0.68$ . The red dotted, green dot-dashed, blue dashed and black solid lines respectively correspond to the case of $\beta=0.5, 0.1, 0.05$ and $0.02$.}
  \label{fig:jeans}
\end{figure}

Obviously, a large $k^2_J$ can be obtained either when the value of $|1+8w_u-3w^2_u-6c^2_{s,ad}|$ is large enough, or when the effective sound speed is adequately small (that is, $c^2_{s,eff}\rightarrow0$). The discussion of effective sound speed will be concretely studied in the next section. Here, we just focus on the squared Jeans wave number by the absolute term for the model and investigate its behaviour around the transition. The evolutional curves of the absolute term are plotted in Fig. \ref{fig:jeans} when $\beta$ is varied. When the parameter $\beta$ is sufficiently small, one can obtain a large enough $k^2_J$ which can satisfy the condition $k^2_J\gg k^2$. To some extent, when the fast transition of the model is obvious enough, one can obtain a relatively viable UDF model.

We have analysed the feasibility of fast transition in the UDF model from the aspect of theory. Whether or not this phenomenon is supported by the cosmic observations, in the next section, we will perform a global fitting for the UDF model with entropic perturbations by the cosmic observational data sets. According to the publicly available \textbf{CosmoMC} package \cite{ref:cosmomc-Lewis2002} and the dark fluid perturbations in the \textbf{CAMB} \cite{ref:camb} code, based on Eqs. (\ref{eq:enpereq1}) and (\ref{eq:enpereq2}), we modified the perturbation equations and added new parameters in the code. In our calculations, we assume the shear perturbation $\sigma_u=0$ and the adiabatic initial conditions will be taken.

\section{Constraint method and results}

\subsection{Implications on CMB anisotropy for the model parameters $\beta$, $a_t$ and $c^2_{eff}$}

\begin{figure}[!htbp]
\includegraphics[width=13cm,height=9cm]{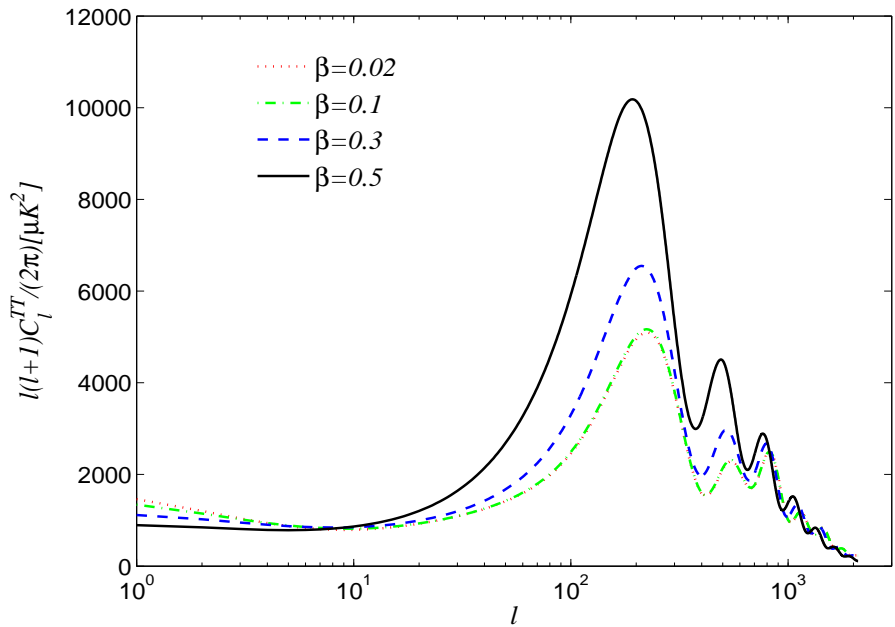}
  \caption{The effects on CMB temperature power spectra of the model parameter $\beta$. The dotted red, green dot-dashed, blue dashed and solid black lines are for $\beta=0.02,0.1,0.3,0.5$ respectively, the other relevant parameters are fixed with the mean values as shown in Table \ref{tab:results}.}
  \label{fig:power-beta}
\end{figure}

\begin{figure}[!htbp]
\includegraphics[width=12cm,height=8cm]{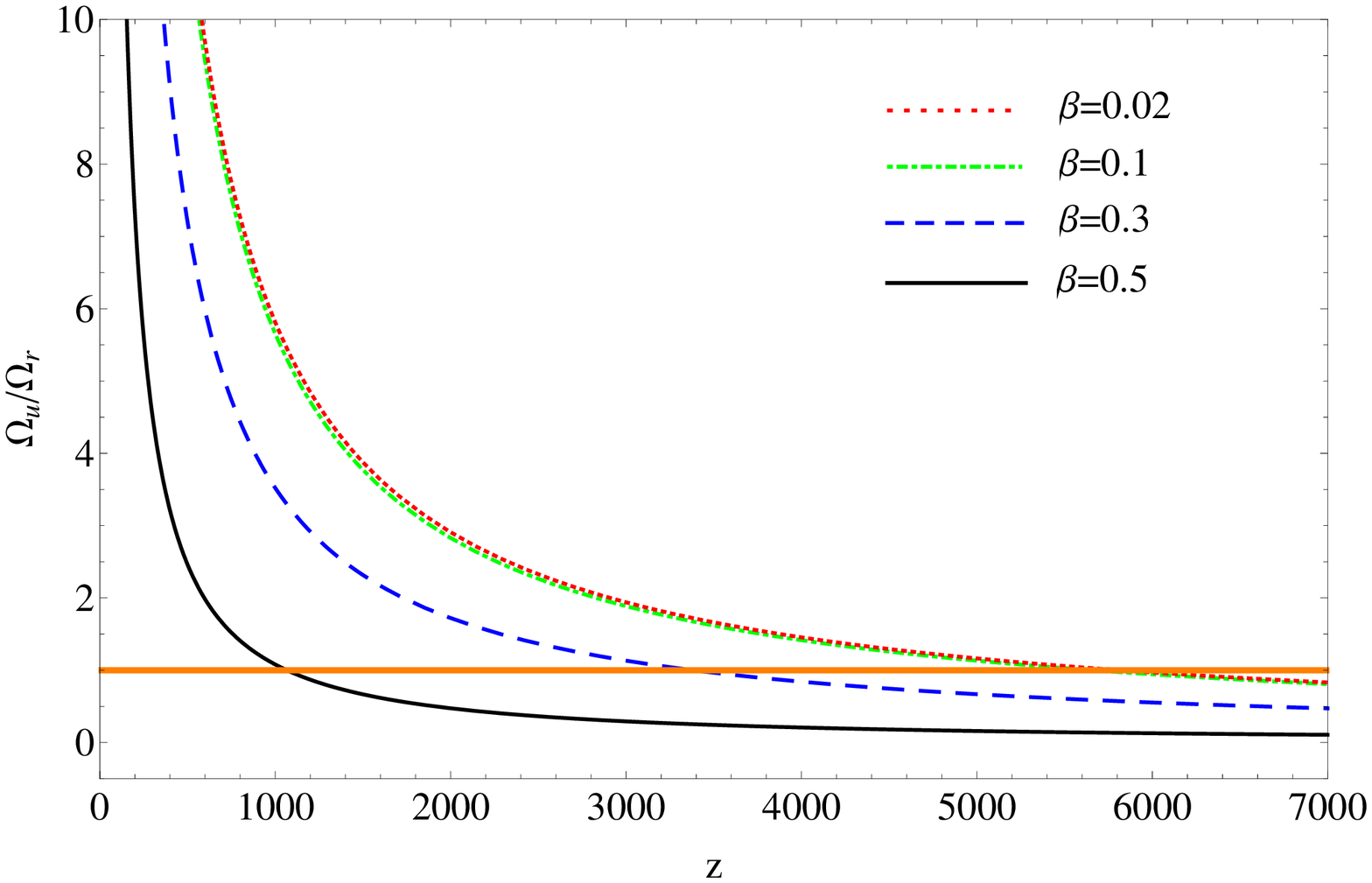}
  \caption{The evolutional curves for the radio of dark fluid and radiation $\Omega_u/\Omega_r$ when the parameter $\beta$ is varied. The dotted red, green dot-dashed, blue dashed and solid black lines are for $\beta=0.02,0.1,0.3,0.5$ respectively, the horizontal orange thick line responds to the case of $\Omega_u=\Omega_r$, the other relevant parameters are fixed with the mean values as shown in Table \ref{tab:results}. In the early epoch, $\Omega_u=\Omega_{dm}$.}
  \label{fig:om-difbeta}
\end{figure}

\begin{figure}[!htbp]
\includegraphics[width=13cm,height=9cm]{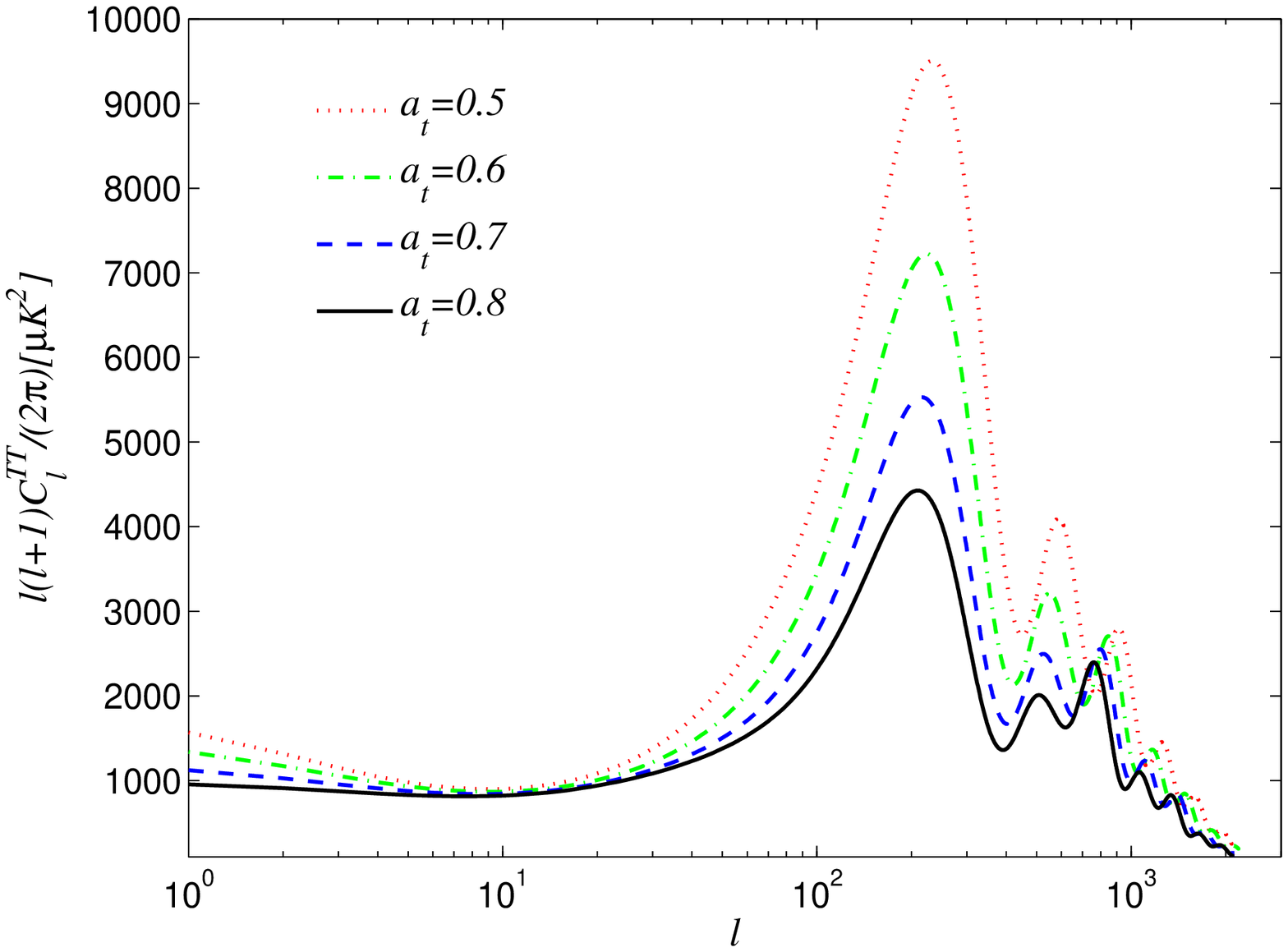}
  \caption{The effects on CMB temperature power spectra of the model parameter $a_t$. The dotted red, green dot-dashed, blue dashed and solid black lines are for $a_t=0.5,0.6,0.7,0.8$ respectively, the other relevant parameters are fixed with the mean values as shown in Table \ref{tab:results}.}
  \label{fig:power-at}
\end{figure}

\begin{figure}[!htbp]
\includegraphics[width=12cm,height=8cm]{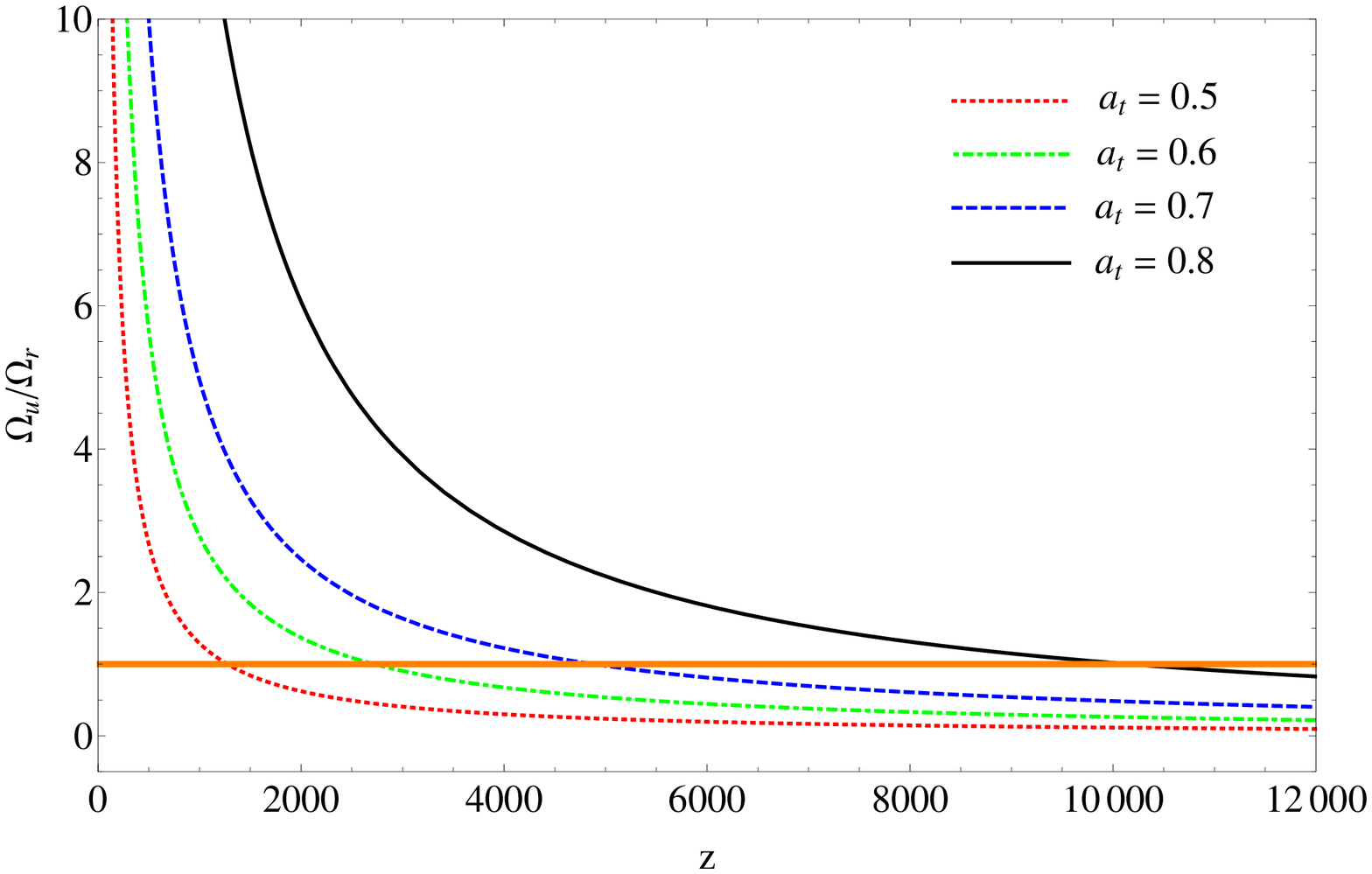}
  \caption{The evolutional curves for the radio of dark fluid and radiation $\Omega_u/\Omega_r$ when the parameter $a_t$ is varied. The dotted red, green dot-dashed, blue dashed and solid black lines are for $a_t=0.5,0.6,0.7,0.8$ respectively, the horizontal orange thick line responds to the case of $\Omega_u=\Omega_r$, the other relevant parameters are fixed with the mean values as shown in Table \ref{tab:results}. In the early epoch, $\Omega_u=\Omega_{dm}$.}
  \label{fig:om-difat}
\end{figure}

To illustrate how the CMB temperature anisotropies are characterized by different values of the model parameters $\beta$ and $a_t$, we respectively plot the effects on the CMB temperature power spectra of $\beta$ and $a_t$ when the other parameters are fixed according to Table \ref{tab:results}. The variation of $\beta$ and $a_t$ will change the value of effective dimensionless energy density of dark matter $\Omega_{dm}$. According to Eq. (\ref{eq:Friedmann}), when the parameter $\beta$ or $a_t$ varies, one can plot the evolutional curves for the radio of dark fluid and radiation $\Omega_u/\Omega_r$, which $\Omega_u=\Omega_{dm}$ in the early epoch. From the Fig. \ref{fig:power-beta} and Fig. \ref{fig:om-difbeta}, decreasing the values of $\beta$ (that is, the transition of EoS becomes faster), which is equivalent to increase the value of effective dimensionless energy density of dark matter $\Omega_{dm}$, will make the equality of matter and radiation earlier, therefore the sound horizon is decreased. As a result, the first peak of CMB temperature power spectra is depressed. Particularly, when $\beta$ = 0.1 and 0.02, the power spectra varies weakly because the value of $\Omega_{dm}$ changes slightly in this two case. Similarly, based on Fig. \ref{fig:power-at} and Fig. \ref{fig:om-difat}, with the increasement of the values of $a_t$, the first peak of $C^{TT}_l$ is depressed.

\begin{figure}[!htbp]
\includegraphics[width=13cm,height=9cm]{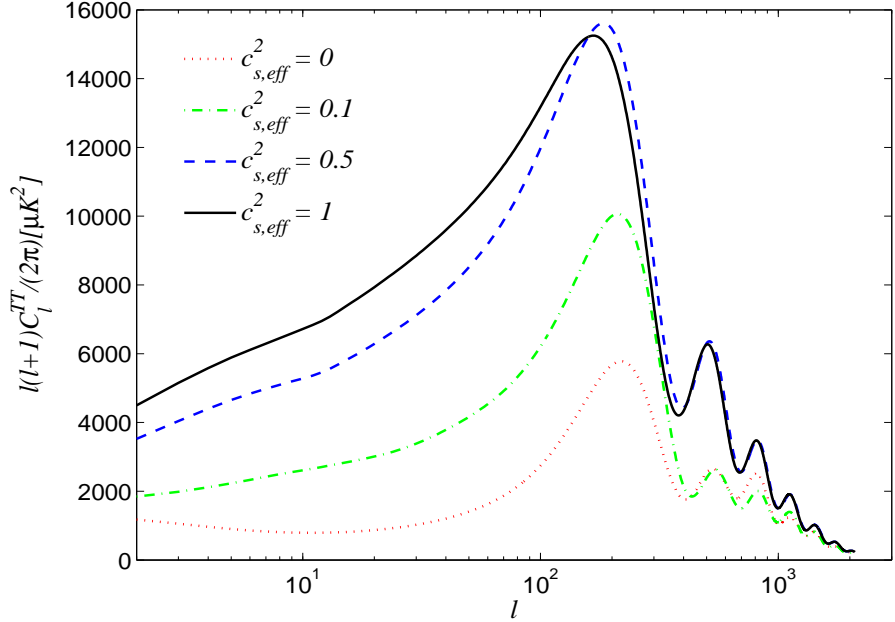}
  \caption{The effects on CMB temperature power spectra of the model parameter $c^2_{s,eff}$. The dotted red, green dot-dashed, blue dashed and solid black lines are for $c^2_{s,eff}=0,0.1,0.5,1$ respectively, the other relevant parameters are fixed with the mean values as shown in Table \ref{tab:results}.}
  \label{fig:power-cs2}
\end{figure}

In addition, When the entropic perturbations are considered, the effects on CMB temperature power spectra of the parameter $c^2_{s,eff}$ are also important. We choose several different values of $c^2_{s,eff}$ in the range $[0,1]$ with the other parameters fixed according to Table \ref{tab:results}. In Fig. \ref{fig:power-cs2}, one can see that when the value of $c^2_{s,eff}$ is large, the gravitational potential decays quickly due to pressure support of the UDF fluctuations during UDF domination. With the decreasement of $c^2_{s,eff}$, the effect on CMB temperature power spectra becomes weak, the case of $c^2_{s,eff}=0$ was analyzed in Ref. \cite{ref:cs2define-Hu1998}. It is clear that CMB power spectra favors small values of the effective sound speed.

\subsection{Constraint method and data sets}
By using the MCMC method, we constrain the following 8-dimensional parameter space for the UDF model
\begin{eqnarray}
P\equiv\{{\omega_b,\Theta_S,\tau,a_t,\beta,c^2_{s,eff},n_s,log[10^{10}A_s]}\},
\label{eq:params}
\end{eqnarray}
where $\omega_b=\Omega_bh^2$ stands for the baryon matter density, $\Theta_S$ refers to the ratio of sound horizon and angular diameter distance, $\tau$ indicates the optical depth, $a_t$ and $\beta$ are the added parameters for the UDF model, $c^2_{s,eff}$ is the effective speed sound, $n_s$ is the scalar spectral index, and $A_s$ represents the amplitude of the initial power spectrum. In Table \ref{tab:priors}, we show the priors of the model parameters.
\begin{table}
\begin{center}
\begin{tabular}{cc}
\hline\hline Parameters&Priors\\ \hline
$\Omega_b h^2$ & $[0.005,1]$ \\
$\Theta_S$ & $[0.5,10]$ \\
$\tau$ & $[0.01,0.8]$ \\
$a_t$ & $[0,1]$ \\
$\beta$ & $[0,1]$ \\
$c^2_{s,eff}$ & $[0,1]$ \\
$n_s$ & $[0.5,1.5]$ \\
$log[10^{10} A_s]$ & $[2.7,4]$ \\
$Age/Gyr$ & $10Gyr<t_0<20Gyr$ \\
$\omega_b$ & $0.022\pm0.002$ \cite{ref:omegab-Burles2001} \\
$H_0$ & $74.2\pm3.6 kms^{-1}Mpc^{-1}$ \cite{ref:H0-Riess2009} \\
\hline\hline
\end{tabular}
\caption{The priors of the model parameters and other priors. $k_{s0}=0.05Mpc^{-1}$ was adopted as the pivot scale of the initial scalar power spectrum.}\label{tab:priors}
\end{center}
\end{table}

In our numerical calculations, the total likelihood $\chi^2$ can be constructed as
\begin{eqnarray}
\chi^2=\chi^2_{CMB}+\chi^2_{BAO}+\chi^2_{SNIa}.
\label{eq:ki2}
\end{eqnarray}

The CMB, BAO and SNIa data points are respectively from WMAP 7-yr data \cite{ref:WMAP7-Komatsu2011}, SDSS data \cite{ref:BAO-Percival2010} and SNIa Union 2.1 data \cite{ref:SN580-Suzuki2012}. More detailed descriptions about the cosmic observations, one can see Ref. \cite{ref:details-xu2010}.

\subsection{Fitting results and discussions}
\begin{figure}[!htbp]
\includegraphics[width=20cm,height=14cm]{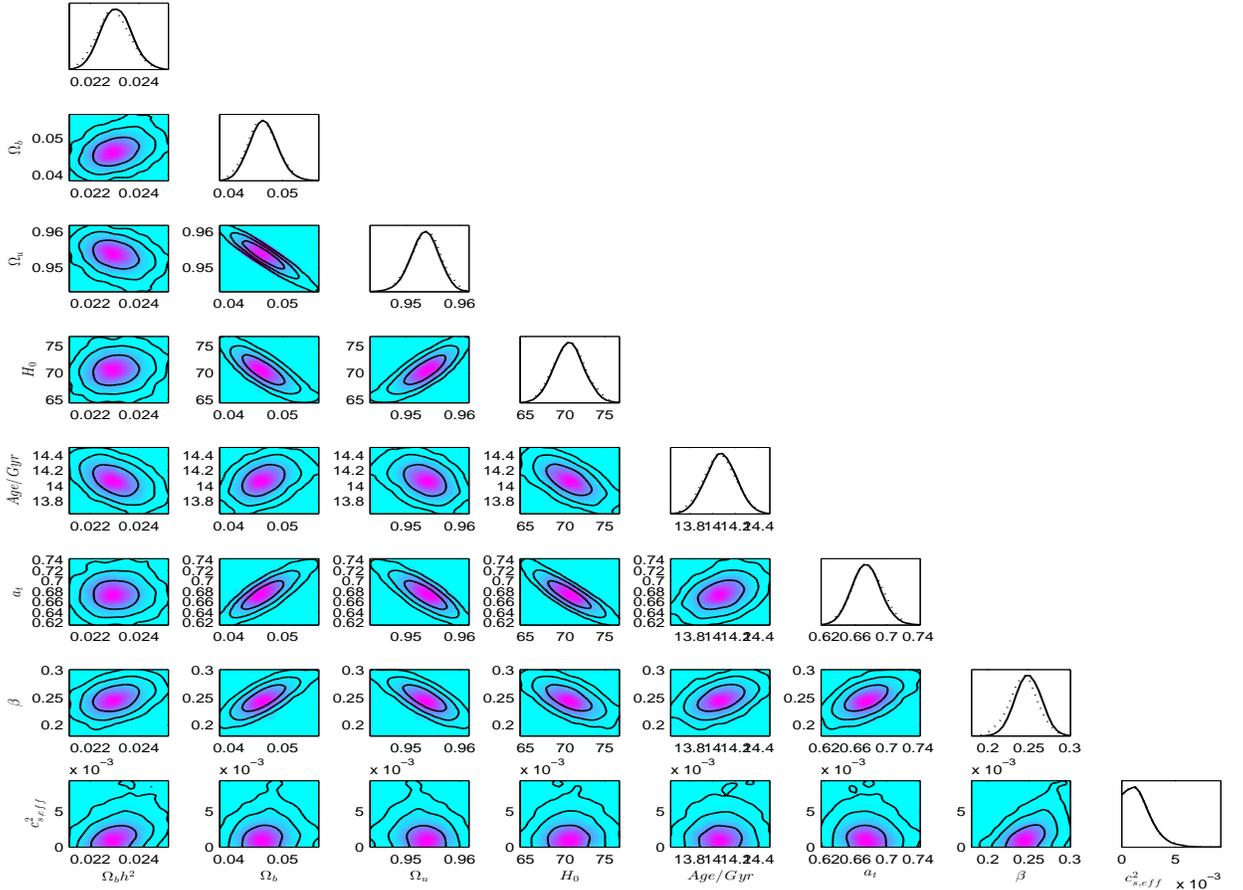}
  \caption{The 1-D marginalized distributions on individual parameters and 2-D contours with 68\%C.L., 95 \%C.L., and 99.7\%C.L. between each other using the combination of the observational data points from the full CMB information, BAO and SNIa for the UDF model.}
  \label{fig:contour}
\end{figure}

We have run 8 chains in parallel on the \textit{Computational Cluster for Cosmos} (3C) and checked the convergence (R-1 is of the order 0.01), the constraint results from observational data sets are presented in Table \ref{tab:results} and Fig. \ref{fig:contour}. In Table \ref{tab:results}, we list the mean values of basic and derived model parameters with 1,2,3$\sigma$ regions. Then, in Fig. \ref{fig:contour}, we show the one-dimensional (1D) marginalized distributions of parameters and 2D contours with confidence level.

\begin{figure}[!htbp]
\includegraphics[width=12cm,height=8cm]{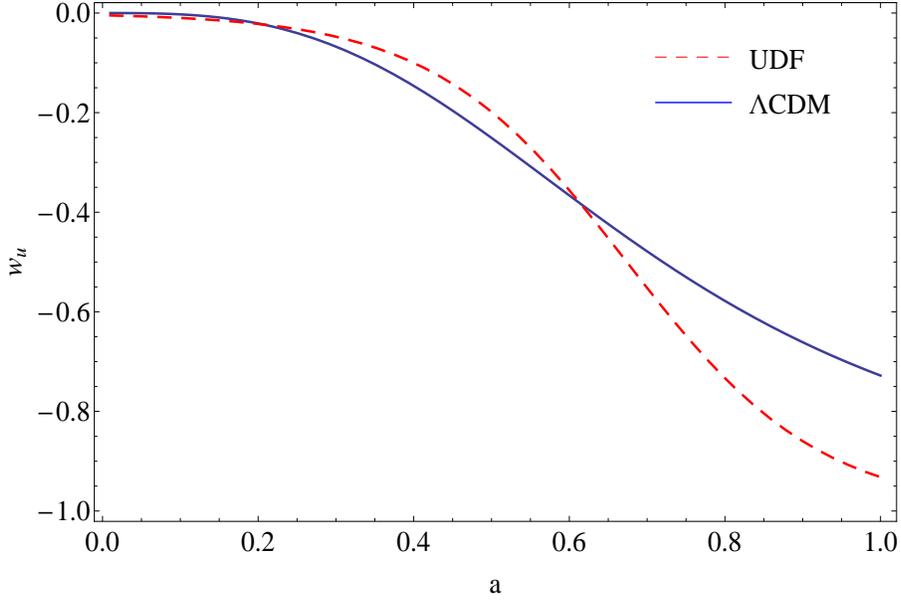}
  \caption{The evolutional curves of EoS for the UDF model and one for $\Lambda$CDM model, where $w_u=\rho_{\Lambda}/(\rho_{\Lambda}+\rho_{CDM})$ for $\Lambda$CDM. In this figure, the red dashed line is for the UDF model with mean values as shown in Table \ref{tab:results}, the blue solid line is for $\Lambda$CDM model with mean values taken from Ref. \cite{ref:WMAP7-Komatsu2011} with WMAP+BAO+$H_0$ constraint results.}
  \label{fig:wvs}
\end{figure}

\begin{figure}[!htbp]
\includegraphics[width=13cm,height=9cm]{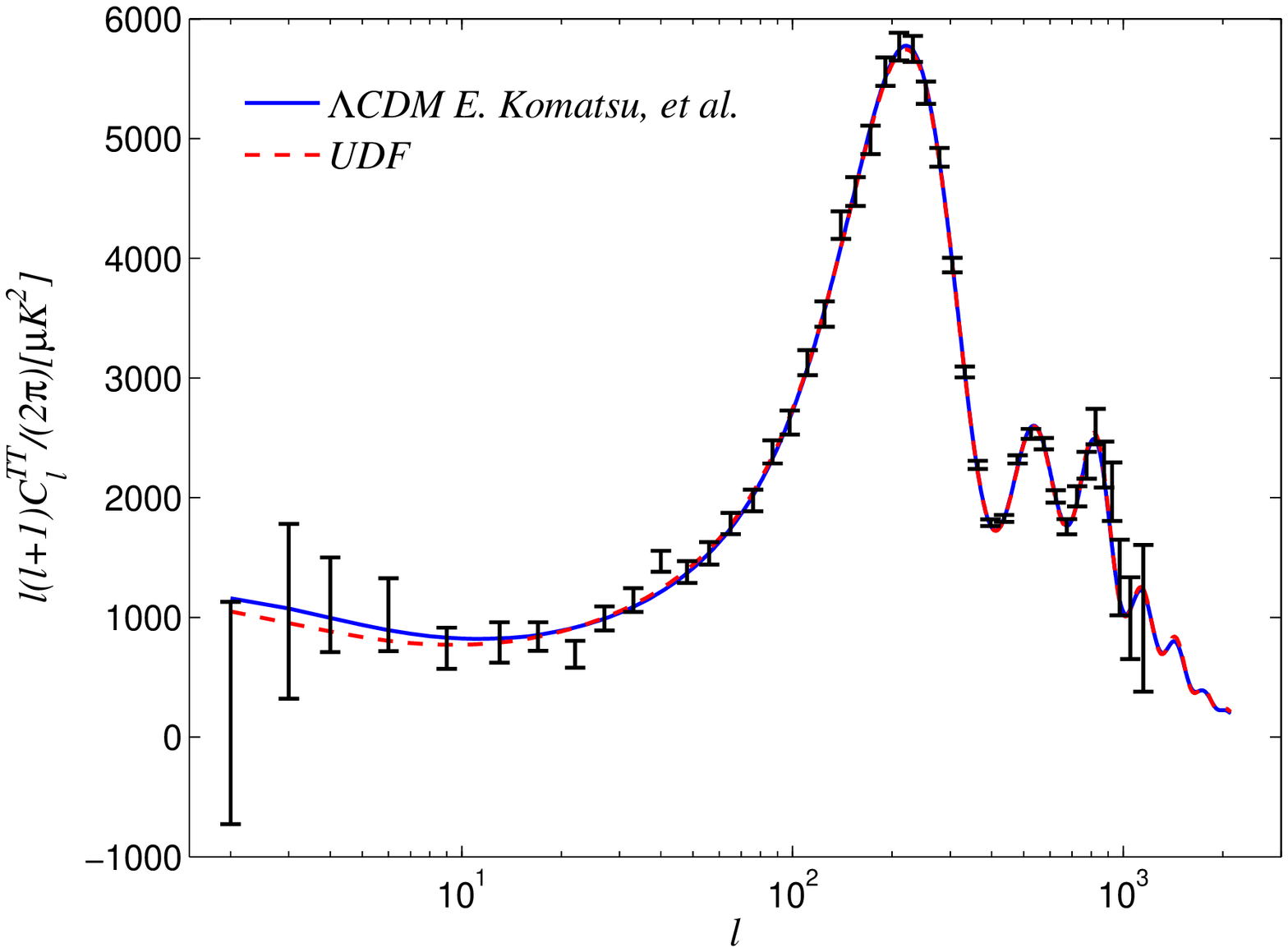}
  \caption{The CMB $C^{TT}_l$ temperature power spectra v.s. multipole moment $l$, where the black dots with error bars denote the observational data points with their corresponding uncertainties from WMAP 7-yr results, the red dashed line is for the UDF model with mean values as shown in Table \ref{tab:results}, the blue solid line is for $\Lambda$CDM model with mean values taken from \cite{ref:WMAP7-Komatsu2011} with WMAP+BAO+$H_0$ constraint results.}
  \label{fig:power}
\end{figure}

The results show that the cosmic data sets from SN, BAO and CMB can give a tight constraint to the model parameter space and favor a model with the scale factor corresponding the moment of transition $a_t = 0.674_{-0.0177}^{+0.0173}$ and the rapidity of the transition $\beta = 0.249_{-0.00714 }^{+ 0.00727}$ in 1$\sigma$ region. By using the obtained mean values and Ref. \cite{ref:WMAP7-Komatsu2011}, we plotted the contrastive evolutional curves of EoS for the UDF and $\Lambda$CDM in Fig. \ref{fig:wvs}. From the figure, one can read off that the UDF behaves like cold dark matter at the early epoch and like dark energy at the late epoch. According to our definition of slow and fast transitions on the small interval $[0.67,0.68]$, the evolution of EoS undergoes a slow transition, which means that the cosmic observations can not favor the phenomenon of fast transition and the differences between the UDF model and the $\Lambda$CDM model are not obvious. Besides, the results also give a tight constraint to the effective sound speed, that is, $c^2_{s,eff} = 0.00168_{- 0.00168}^{+0.000347}$ in 1$\sigma$ region. When the entropy perturbation is included, a negative adiabatic sound speed is favored, which is different from that of the pure adiabatic case. Particularly, although the transition is not fast, the small enough value of effective sound speed $c^2_{s,eff}$ can also make the UDF model relatively viable because of $k^2_J\gg k^2$, which is based on the analysis about Eq. (\ref{eq:jeans}). We also show the $C^{TT}_l$ power spectra for the $\Lambda$CDM model \cite{ref:WMAP7-Komatsu2011} and observational data points in Fig. \ref{fig:power} where the mean values of relevant model parameters are adopted. It implies that the cosmic observational data points cannot discriminate the $\Lambda$CDM model from the UDF model.

\begin{table}
\begin{center}
\begin{tabular}{cc}
\hline\hline Parameters&Mean with errors\\ \hline
$\Omega_b h^2$ & $    0.0231_{-    0.000613-    0.00116-    0.00170}^{+    0.000603+    0.00126+    0.00189}$ \\
$\Omega_b$ & $    0.0465_{-    0.00252-    0.00477-    0.00724}^{+    0.00247+    0.00526+    0.00900}$ \\
$\Omega_u$ & $    0.953_{-    0.00247-    0.00526-    0.00898}^{+    0.00252+    0.00477+    0.00725}$ \\
$H_0$ & $   70.486_{-    1.810-    3.568-    5.406}^{+    1.779+    3.581+    5.662}$ \\
$Age/Gyr$ & $   14.074_{-    0.130-    0.251-    0.380}^{+    0.128+    0.251+    0.375}$ \\
$a_t$ & $    0.674_{-    0.0177-    0.0341-    0.0508}^{+    0.0173+    0.0369+    0.0612}$ \\
$\beta$ & $    0.249_{-    0.00714-    0.0249-    0.0413}^{+    0.00727+    0.0247+    0.0424}$ \\
$c^2_{s,eff}$ & $    0.00168_{-    0.00168-    0.00168-    0.00168}^{+    0.000347+    0.00224+    0.00521}$ \\
\hline\hline
\end{tabular}
\caption{The constraint results of basic and derived model parameters with $1,2,3\sigma$ regions from the full CMB information, BAO and SNIa data sets.}
\label{tab:results}
\end{center}
\end{table}

In order to further judge the difference between the UDF model and the $\Lambda$CDM model from the objective model selection criteria, the Akaike information criterion (AIC) \cite{ref:AIC-define} is employed. The definition is
\begin{eqnarray}
AIC = -2ln\mathcal {L} (\theta\mid data)_{max}+ 2\mathcal {P},
\label{eq:AIC-define}
\end{eqnarray}
where $\mathcal {L}_{max}$ is the highest likelihood in the model with the best fit parameters $\theta$ and $-2ln\mathcal {L} (\theta\mid data)_{max}=\chi^2_{min}$, $\mathcal {P}$ is the number of estimable parameter $\theta$ in the model.

Taking the $\Lambda$CDM model as the standard model, the rules for judging the AIC model selection are as follows \cite{ref:AIC-rules}. If $0\leq\Delta AIC\leq2$, the model is regarded as to be equivalent to the $\Lambda$CDM model. When $2\leq\Delta AIC\leq4$, the model is supported considerably less by the standard model. If $\Delta AIC>10$, the model is practically irrelevant with the $\Lambda$CDM model. In this paper, we use the same cosmic observations to constrain the $\Lambda$CDM model. The results show that $\chi^2_{\Lambda CDM}=8024.058$ and $\chi^2_{UDF}=8023.45$, $\Delta AIC=0.608\in(0,2)$, so, the UDF model is equivalent to the $\Lambda$CDM. According to our definition for slow and fast transitions, the $\Lambda$CDM undergoes a slow transition obviously. Therefore, we conclude that the cosmic observations do not favor the phenomenon of fast transition in the UDF model.

\subsection{Some other discussions}
\subsubsection{The variable effective sound speed}

The speed of sound can also be parameterized as a function of the scale factor, such as Refs. \cite{ref:UDF12,ref:fast3-Bertacca2011}. Following Ref. \cite{ref:fast3-Bertacca2011}, the effective sound speed reads
\begin{eqnarray}
c^2_{s,eff} = \frac{c^2_{\infty}}{1+(\nu a^{-3}/c^2_{\infty})^n},
\label{eq:cs2-parameter}
\end{eqnarray}
where $c^2_{\infty}>0$ and $n>0$ are free parameters, $\nu=\Omega_{dm}/\Omega_{de}$ is shown in Ref. \cite{ref:UDF12}. In our UDF model, the dark matter and dark matter do not be decomposed. So $\nu=3/7$ will be taken in our calculations.

\begin{figure}[!htbp]
\includegraphics[width=20cm,height=14cm]{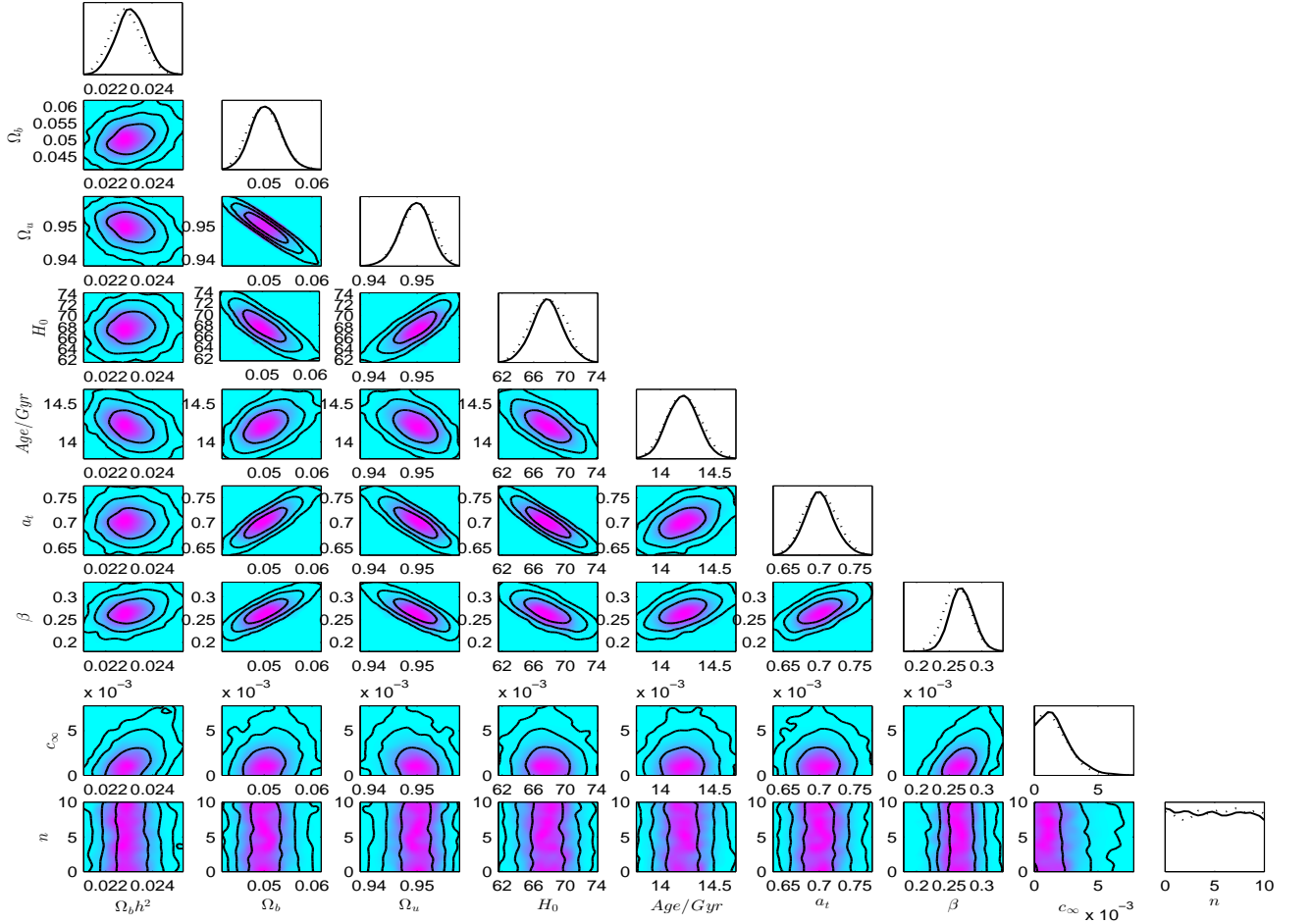}
  \caption{The 1-D marginalized distributions on individual parameters and 2-D contours with 68\%C.L., 95 \%C.L., and 99.7\%C.L. between each other using the combination of the observational data points from the full CMB information, BAO and SNIa, when the effective sound speed $c^2_{s,eff}$ is variable.}
  \label{fig:cs2_contour}
\end{figure}

In the same way, based on the full CMB, BAO and SNIa data ponits, we adopt the MCMC method to constrain the UDF model with the variable $c^2_{s,eff}$. The results are shown in Tab. \ref{tab:results-cs2}. Then, we show the one-dimensional (1D) marginalized distributions of parameters and 2D contours with confidence level in Fig. \ref{fig:cs2_contour}. The constraint for the parameter $n$ is not tight. According to Eq. (\ref{eq:cs2-parameter}), it is easy to conclude that the observations favor a very small value of the effective sound speed. The observational data points do not favor the phenomenon of fast transition in the UDF model.

\begin{table}
\begin{center}
\begin{tabular}{cc}
\hline\hline Prameters&Mean with errors\\ \hline
$\Omega_b h^2$ & $    0.0231_{-    0.000617-    0.00119-    0.00175}^{+    0.000610+    0.00122+    0.00190}$ \\
$\Omega_b$ & $    0.0504_{-    0.00296-    0.00568-    0.00796}^{+    0.00296+    0.00608+    0.00987}$ \\
$\Omega_u$ & $    0.950_{-    0.00296-    0.00608-    0.00982}^{+    0.00296+    0.00568+    0.00797}$ \\
$H_0$ & $   67.759_{-    1.857-    3.734-    5.564}^{+    1.872+    3.916+    5.687}$ \\
$Age/Gyr$ & $   14.207_{-    0.138-    0.268-    0.384}^{+    0.139+    0.280+    0.429}$ \\
$a_t$ & $    0.701_{-    0.0197-    0.0383-    0.0540}^{+    0.0196+    0.0419+    0.0636}$ \\
$\beta$ & $    0.269_{-    0.00834-    0.0280-    0.0492}^{+    0.00769+    0.0286+    0.0515}$ \\
$c^2_{\infty}$ & $    0.00177_{-    0.00177-    0.00177-    0.00177}^{+    0.000356+    0.00242+    0.00477}$ \\
$n$ & $    4.931_{-    4.927-    4.927-    4.927}^{+    3.428+    4.750+    5.051}$ \\
\hline\hline
\end{tabular}
\caption{The constraint results of basic and derived model parameters for the UDF model with a variable $c^2_{s,eff}$ with $1,2,3\sigma$ regions from the full CMB, BAO and SNIa data points.}
\label{tab:results-cs2}
\end{center}
\end{table}

\subsubsection{Comparison between the CMB shift parameter and the full CMB information}

In order to make sure that the full CMB information can give a tighter constraint than the CMB shift parameter R (CMBR) covariance matrix \cite{ref:fast4-Felice2012,ref:cs2studyfree-Xu2012}, we will constrain the UDF model by using the CMBR, BAO and SNIa data points. The unified fluid have to be decomposed as dark matter and dark energy, $\rho_u=\rho_{dm}+\rho_{de}$, when the CMBR data is used. The conservation equation of the UDF is
\begin{eqnarray}
\rho_u+3H(1+w_u)\rho_u=0.
\label{eq:contotal}
\end{eqnarray}

By introducing a intermediate variable $\Gamma$, the above equation becomes
\begin{eqnarray}
\dot{\rho_{dm}}+3H(1+w_u)\rho_{dm}=-\Gamma,
\label{eq:condetach1}
\end{eqnarray}
\begin{eqnarray}
\dot{\rho_{de}}+3H(1+w_u)\rho_{de}=\Gamma.
\label{eq:condetach2}
\end{eqnarray}

If the dark matter is conserved, one can obtain the evolutional equation of energy density for the dark energy
\begin{eqnarray}
\frac{d\rho_{de}}{dlna}=-3[(1+w_u)\rho_{de}+w_u\rho_{dm}].
\label{eq:rhode}
\end{eqnarray}

\begin{figure}[!htbp]
\includegraphics[width=20cm,height=14cm]{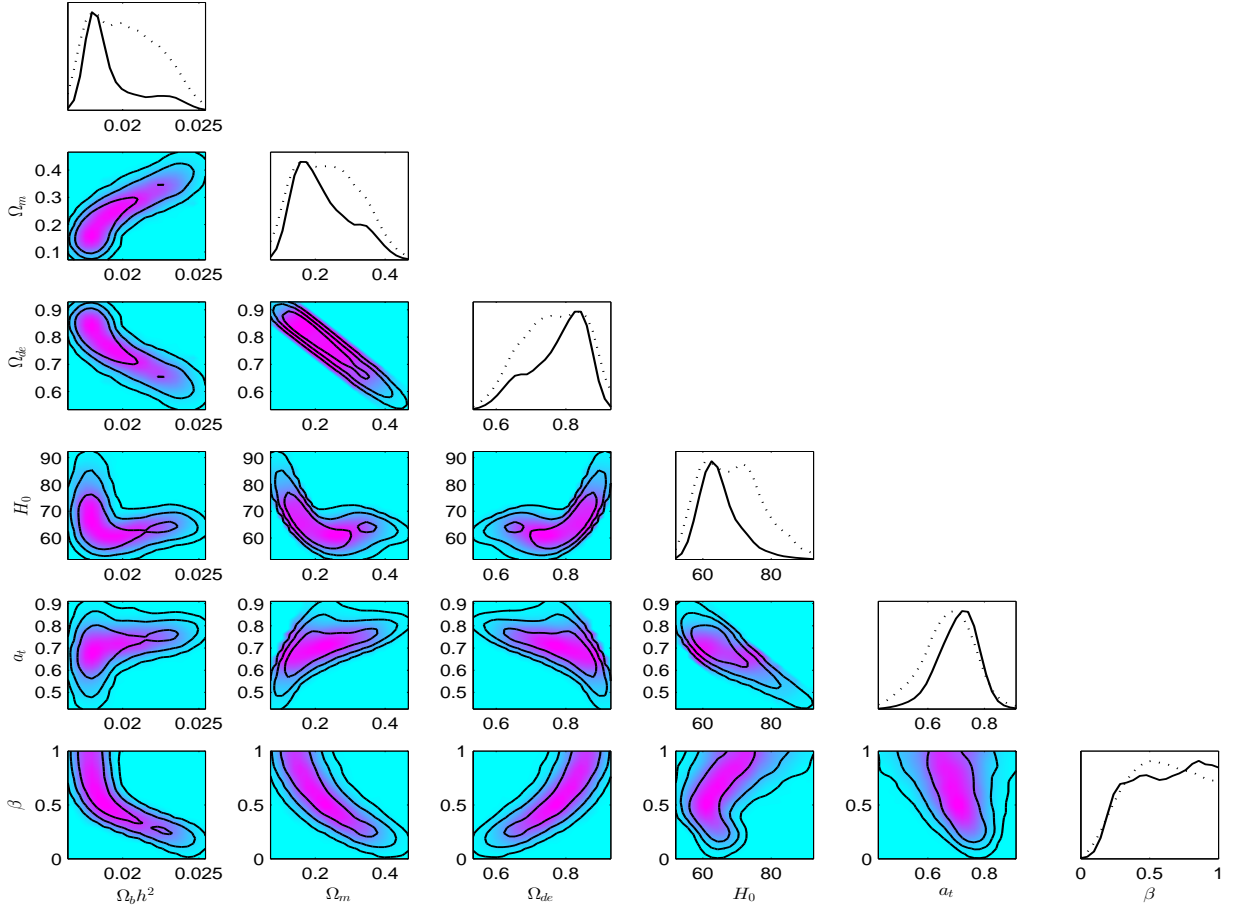}
  \caption{The 1-D marginalized distributions on individual parameters and 2-D contours with 68\%C.L., 95 \%C.L., and 99.7\%C.L. between each other using the combination of the observational data points from CMBR, BAO and SNIa, when the effective sound speed $c^2_{s,eff}$ is a constant.}
  \label{fig:wcmbr}
\end{figure}

Then we modify the \textbf{CAMB} package \cite{ref:camb} about the energy density of the dark matter and dark energy. The constraint mean values of model parameters are shown in Table \ref{tab:results_wCMBR} and the one-dimensional (1D) marginalized distributions of parameters and 2D contours are plotted in Fig. \ref{fig:wcmbr}. The results tell us the observational data points do not favor the phenomenon of fast transition, however, the constraint for some model parameters are not tight and reliable. The constraint for the parameter $\beta$ is not tight. The 1-D marginalized distributions for the model parameters are apparently skewness. According to Ref. \cite{ref:cosmomc-Lewis2002}, the dotted lines represent mean likelihoods of samples and the solid lines are marginalized probabilities. For gaussian distributions, they are the same. For skewed distributions, they are different. Of course, the gaussian distributions fit better than the skewed ones. A possible way to make the distributions gaussian is to increase the observational data points. So the full CMB information will be a more reasonable choice than the CMBR data.

\begin{table}
\begin{center}
\begin{tabular}{cc}
\hline\hline Prameters&Mean with errors\\ \hline
$\Omega_b h^2$ & $    0.0193_{-    0.00148-    0.00199-    0.00241}^{+    0.00215+    0.00462+    0.00573}$ \\
$\Omega_{m}$ & $    0.222_{-    0.0776-    0.111-    0.133}^{+    0.0939+    0.169+    0.222}$ \\
$\Omega_{de}$ & $    0.778_{-    0.0939-    0.169-    0.222}^{+    0.0776+    0.111+    0.134}$ \\
$H_0$ & $   65.007_{-    5.246-    8.469-   10.996}^{+    5.506+   14.734+   25.052}$ \\
$a_t$ & $    0.704_{-    0.0675-    0.158-    0.250}^{+    0.0651+    0.117+    0.181}$ \\
$\beta$ & $    0.606_{-    0.144-    0.383-    0.485}^{+    0.394+    0.394+    0.394}$ \\
\hline\hline
\end{tabular}
\caption{The constraint results of basic and derived model parameters for the UDF model with a invariable $c^2_{s,eff}$ with $1,2,3\sigma$ regions from CMBR, BAO and SNIa data points.}
\label{tab:results_wCMBR}
\end{center}
\end{table}

\section{Summary}

Inspired by the previous papers \cite{ref:fast1-Bruni2012,ref:cs2studyfree-Xu2012}, we have proposed a UDF model with fast transition between an early matter dominated era and a late $\Lambda$CDM-like epoch. In particular, to avoid the instability of the perturbations when the EoS of the model is negative, the entropic perturbations are considered by introducing the effective sound speed. Then, we have studied the viability of the unified model by the analytical expression of the squared Jeans wave number $k^2_J$ from the aspect of theory. The UDF model can present a large enough squared wave number $k^2_J$ when the fast transition happens, that is, the condition $k^2_J\gg k^2$ can be satisfied. Additionally, We also have studied the effects on CMB temperature power spectra of the model parameters $c^2_{s,eff}$, $a_t$ and $\beta$.

Importantly, by using MCMC method, a global fitting has been performed on the UDF model with the entropic perturbations from SNIa, BAO and the full CMB data sets. All the cosmological parameters for the model are well constrained, which is shown in Table \ref{tab:results}. In addition, we also have constrained the UDF model with a variable $c^2_{s,eff}$ and made a comparison between the full CMB information and the CMBR data. The analysis tells us that the unified fluid behaves like cold dark matter at the early epoch and like dark energy at the late era. We conclude that the phenomenon of fast transition for the UDF model is not favored by the cosmic observations, the evolution of the EoS undergoes a slow transition and the evolutional curves of CMB power spectra for the UDF and $\Lambda$CDM are highly similar, so the differences with $\Lambda$CDM are negligible. The same conclusion can be obtained when we use the Akaike information criterion. Besides, the cosmic observations favor a small amount of entropic perturbations, the pure adiabatic perturbations can not be ruled out. Meanwhile, the value of effective sound speed $c^2_{s,eff}$ is small enough which can also make the UDF model relatively viable. We look forward to more cosmic observations which can further constrain the UDF models and make us acquaint with this kind of model clearly.

\acknowledgements{L. Xu's work is supported in part by NSFC under the Grants No. 11275035 and "the Fundamental Research Funds for the Central Universities" under the Grants No. DUT13LK01.}

\end{document}